\documentstyle{article}
\input amssym.def
\input amssym
\def\ol{\overline}
\def\tr{\mbox{Tr}}
\def\tp#1{#1^{\mbox{\scriptsize{T}}}}
\def\dim{\mbox{dim}\;}
\def\a{\alpha}
\def\b{\beta}
\def\c{\gamma}
\def\d{\delta}
\def\s{\sigma}
\def\w{\omega}
\def\W{\Omega}
\def\fa{\frak A}
\def\fb{\frak B}
\def\fc{\frak C}
\def\dg{\dagger}
\def\ds{\displaystyle}
\def\lra{\Leftrightarrow}
\def\dirsum#1{\mbox{{\normalsize ${\ds\bigoplus_{#1}}$}}}
\def\ix{{\rm ad,F,B}}
\def\mb#1{\mbox{\boldmath{$#1$}}}
\newtheorem{defn}{Definition}
\newtheorem{prop}{Proposition}
\newtheorem{rem}{Remark}
\newtheorem{thm}{Theorem}
\newcommand{\be}{\begin{equation}}
\newcommand{\ee}{\end{equation}}
\newcommand{\bea}{\begin{eqnarray}}
\newcommand{\eea}{\end{eqnarray}}
\newcommand{\beax}{\begin{eqnarray*}}
\newcommand{\eeax}{\end{eqnarray*}}
\newcommand{\no}{\noindent}
\newcommand{\op}[1]{\mbox{\sf #1}}
\newcommand{\komm}[2]{\left[#1,#2\right]}
\newcommand{\akomm}[2]{\left\{#1,#2\right\}}

\newcommand{\bd}{\begin{defn}{\hspace{-.55em}\em{\bf {: }}}}
\newcommand{\ed}{\end{defn}}
\newcommand{\bp}{\begin{prop}{\hspace{-.55em}\em{\bf {: }}}}
\newcommand{\ep}{\end{prop}}
\newcommand{\br}{\begin{rem}{\hspace{-.55em}\em{\bf {: }}}}
\newcommand{\er}{\end{rem}}
\newcommand{\bt}{\begin{thm}{\hspace{-.55em}\em{\bf {: }}}}
\newcommand{\et}{\end{thm}}

\begin{document}
\bibliographystyle{plain}
\begin{titlepage}
\renewcommand{\thefootnote}{\fnsymbol{footnote}}
\large
\hfill\begin{tabular}{l}HEPHY-PUB 654/96\\ UWThPh-1996-50\\ hep-th/9711176\\
November 1997
\end{tabular}\\[3cm]
\begin{center}
{\Large\bf CLIFFORD ALGEBRAS IN}\\[.5ex]
{\Large\bf FINITE QUANTUM FIELD THEORIES}\\[1cm]
{\Large\bf II.~Reducible Yukawa Finiteness Condition}\\
\vspace{1.7cm}
{\Large\bf Wolfgang LUCHA\footnote[1]{\normalsize\ E-mail:
v2032dac@awiuni11.edvz.univie.ac.at}}\\[.5cm]
Institut f\"ur Hochenergiephysik,\\
\"Osterreichische Akademie der Wissenschaften,\\
Nikolsdorfergasse 18, A-1050 Wien, Austria\\[1.7cm]
{\Large\bf Michael MOSER\footnote[2]{\normalsize\ E-mail:
mmoser@galileo.thp.univie.ac.at}}\\[.5cm]
Institut f\"ur Theoretische Physik,\\
Universit\"at Wien,\\
Boltzmanngasse 5, A-1090 Wien, Austria\\[2cm]
{\bf Abstract}
\end{center}
\normalsize
An arbitrary renormalizable quantum field theory is considered as finite if
its dimensionless couplings conspire to yield at every order of its
perturbative expansion no ultraviolet-divergent renormalizations of the
physical parameters of the theory. The ``finiteness conditions'' resulting
from these requirements form highly complicated, non-linear systems of
relations. A promising type of solution to the condition for one-loop
finiteness of the Yukawa couplings involves Yukawa couplings which are
equivalent to~the generators of Clifford algebras with identity element.
However, our attempt to construct a finite model based on such Clifford-like
Yukawa couplings fails: a Clifford structure of the Yukawa couplings spoils
finiteness of the gauge couplings, at least for every simple gauge group of
rank less than or equal~to~8.

\vspace*{6ex}

\normalsize
\renewcommand{\thefootnote}{\arabic{footnote}}
\end{titlepage}

\section{Introduction}

In a recent paper \cite{lucha96-1}, henceforth called Part~I, we started the
investigation of the relevance of Yukawa coupling matrices which are
equivalent to the generators (in some representation) of a Clifford algebra
with identity element for the construction of (four-dimensional) finite
quantum field theories. A finite quantum field theory is a renormalizable
quantum field theory which, when evaluated perturbatively, is not burdened by
the appearance of any ultraviolet divergences in the renormalizations of its
physical parameters, i.e., masses and coupling constants, and the necessity
of making sense of this theory by application of the renormalization
programme. Until now, only certain classes of (simple~or~extended)
supersymmetric theories have been safely demonstrated to be finite to all
orders of the loop expansion. (For some brief account of the history of these
developments, see, e.g., Ref.~\cite{lucha96-1} and Sec.~II of
Ref.~\cite{lucha86-1}.)

Soon after the discovery of these supersymmetric finite quantum field
theories, an obvious question has been raised \cite{lucha86-1,lucha86-2}: Is
supersymmetry a {\em necessary\/} ingredient for the construction of finite
quantum field theories? The inspection of general gauge theories immediately
showed that, in order to have the prospect to become finite, a quantum field
theory must contain (interacting) {\em vector bosons, fermions, and scalar
bosons\/} \cite{lucha86-1,lucha86-2,boehm87,lucha87-1}. Moreover, in all
special cases considered, the requirement of the absence of quadratic
divergences seems to point to the supersymmetry of the model under
consideration \cite{qdiv1,qdiv2}.

In a comprehensive study of the conditions for finiteness of a general
quantum field theory reported in Ref.~\cite{kranner91}, it has been argued
that choosing the involved Yukawa coupling matrices to be equivalent~to the
generators of a Clifford algebra with identity element might facilitate the
solution of the finiteness conditions. Picking up this compelling idea, let's
try to construct such a ``Clifford-based'' finite theory.

The outline of this paper is as follows: In Sec.~\ref{sec:fcigqft}, in order
to set the stage for the present discussion, we recall the lowest-order
conditions for finiteness of arbitrary renormalizable quantum field~theories.
Within this system of central finiteness conditions, the analysis of the
condition for one-loop finiteness of Yukawa couplings proves to be already a
rather crucial step in our hunt for additional finite~theories. Considering
the behaviour of the Yukawa coupling matrices under gauge transformations, we
succeed to decompose, in Sec.~\ref{sec:ryfc}, this (one-loop) Yukawa
finiteness condition into its ``irreducible'' components, that is, into
subsystems which are not linked to each other by gauge transformations. The
irreducible subsystems have already been exhaustively discussed in Part~I
\cite{lucha96-1}. Accordingly, we focus our attention in Sec.~\ref{sec:cred1}
to the additional ``degree of freedom'' introduced by this particular kind of
reducibility of~the Yukawa finiteness condition with respect to gauge
transformations. By making excessive use of results obtained in Part~I
\cite{lucha96-1}, it is a rather straightforward task to analyse the full
Yukawa finiteness condition with respect to the above Clifford-algebra
conjecture \cite{kranner91}. We do this in Sec.~\ref{sec:rca} for the
particular class~of candidates for finite theories extracted in
Ref.~\cite{kranner91} and in Sec.~\ref{sec:rcagt} for an arbitrary quantum
field theory. Unfortunately, we are forced to conclude, in
Sec.~\ref{sec:sco}, that the Clifford structure of Yukawa couplings in finite
quantum field theories is incompatible with one- and two-loop finiteness of
the gauge couplings.

\section{Finiteness Conditions in General Quantum Field
Theories}\label{sec:fcigqft}

As for the discussion in Part~I, we consider the most general
\cite{llewellyn73} renormalizable quantum field theory (for particles up to
spin 1 $\hbar$) invariant with respect to gauge transformations forming some
compact simple Lie group $G$ with corresponding Lie algebra $\fa$. The
particle content of this theory is described by
\begin{itemize}
\item (gauge) vector-boson fields $A_\mu(x)=(A_\mu^a)(x)\in\fa$ in the
adjoint representation $R_{{\rm ad}}$: $\fa\rightarrow\fa$ of the gauge group
$G$, of dimension $d_{\rm g}:=\dim\fa$;
\item two-component (Weyl) fermion fields $\psi(x)=(\psi^i)(x)\in V_{\rm F}$
in some representation $R_{\rm F}$: $V_{\rm F}\rightarrow V_{\rm F}$ of $G$,
of dimension $d_{\rm F}:=\dim V_{\rm F}$; and
\item (Hermitean) scalar-boson fields $\phi(x)=(\phi^\a)(x)\in V_{\rm B}$ in
some real representation $R_{\rm B}$: $V_{\rm B}\rightarrow V_{\rm B}$ of
$G$, of dimension $d_{\rm B}:=\dim V_{\rm B}$.
\end{itemize}
Apart from terms involving dimensional parameters, like mass terms and cubic
self-interaction terms of scalar bosons, as well as gauge-fixing and ghost
terms, the Lagrangian defining this theory is given by
\bea
\cal L&=&-\frac{1}{4}\,F_{\mu\nu}^a\,F^{\mu\nu}_a
+i\,\ol{\psi}_i\,\ol{\s}^\mu\left[(D_\mu)_{\rm F}\,\psi\right]^i+\frac{1}{2}
\left[(D_\mu)_{\rm B}\phi\right]^\a\left[(D^\mu)_{\rm B}\phi\right]_\a
\nonumber\\[1ex]
&&-\frac{1}{2}\,\phi^\a\,Y_{\a ij}\,\psi^i\,\psi^j
-\frac{1}{2}\,\phi_\a\,Y^{\dg\a ij}\,\ol{\psi}_i\,\ol{\psi}_j
-\frac{1}{4!}\,V_{\a\b\c\d}\,\phi^\a\,\phi^\b\,\phi^\c\,\phi^\d\ .
\label{fin1}
\eea
Our notations are the same as in Part~I: The Hermitean generators $T_{\rm
R}^a$, ${\rm R} = \ix$, $a=1,2,\dots,d_{\rm g}$, of the Lie algebra ${\fa}$
in the three representations $R_{{\rm ad}}$, $R_{\rm F}$, and $R_{\rm B}$
satisfy the commutation relations
\be
\komm{T_{\rm R}^a}{T_{\rm R}^b}=i\,{f^{ab}}_c\,T_{\rm R}^c\ ,\quad
{\rm R} = \ix\ ,
\label{fin4}
\ee
with the structure constants ${f^{ab}}_c$, $a,b,c=1,2,\dots,d_{\rm g}$,
defining the Lie algebra $\fa$ under consideration. The gauge coupling
constant is denoted by $g$. The gauge-covariant field strength tensor
$F_{\mu\nu}^a$ is of the usual form,
\be
F_{\mu\nu}^a=\partial_\mu A_\nu^a-\partial_\nu A_\mu^a
+g\,{f^a}_{bc}\,A_\mu^b\,A_\nu^c\ .
\label{fin2}
\ee
The gauge-covariant derivatives $D_\mu$ acting on the three representation
spaces $\fa$, $V_{\rm F}$, $V_{\rm B}$, respectively, read
\be
(D_\mu)_{\rm R}:=\partial_\mu-i\,g\,T_{\rm R}^a\,A_\mu^a\ ,\quad
{\rm R} = \ix\ .
\label{fin3}
\ee
The four $2\times 2$ matrices $\ol{\s}^\mu$ in the kinetic term for the Weyl
fermion fields are defined in terms of the $2\times 2$ unit matrix,
$\op{1}_2$, and the three Pauli matrices, $\mb{\s}$, by
$\ol{\s}^\mu=(\op{1}_2,-\mb{\s})$. Without loss of generality, all the Yukawa
couplings $Y_{\a ij}$ may be assumed to be completely symmetric in their
fermionic indices $i$ and $j$, and all the quartic scalar-boson
self-couplings $V_{\a\b\c\d}$ may be taken to be completely symmetric under
an arbitrary permutation of their indices. The group invariants for an
arbitrary representation $R$ of $G$ are defined in terms of the generators
$T_R^a$ of ${\fa}$ in this representation $R$ as usual: the quadratic Casimir
operator $C_R$ is given by
\be
C_R:=\sum_{a=1}^{d_{\rm g}}T_R^a\,T_R^a\ ,
\label{fin5}
\ee
and the Dynkin index $S_R$ is obtained from
\be
S_R\,\delta^{ab}:=\tr\left(T_R^a\,T_R^b\right)\ .
\label{fin6}
\ee
In the adjoint representation $R_{{\rm ad}}$, the Casimir eigenvalue $c_{\rm
g}$ equals the Dynkin index $S_{\rm g}$, i.e., $c_{\rm g}=S_{\rm g}$.

According to our understanding of ``finiteness'' of a general renormalizable
quantum field theory,\footnote{\normalsize\ For a more detailed discussion of
our notion of ``finiteness'' of arbitrary renormalizable quantum field
theories, consult, for instance, Refs.~\cite{lucha96-1,lucha86-1,lucha86-2}.}
finiteness is tantamount to the vanishing of the beta functions of all
physical parameters of this theory, in case of perturbative evaluation of
this quantum field theory order by order in its loop expansion. By
application of the standard renormalization procedure with the help of
dimensional regularization in the minimal-subtraction scheme, the finiteness
conditions relevant here may be easily extracted~\cite{cheng74}, see also
Refs.~\cite{lucha86-1,lucha86-2}:
\begin{itemize}
\item The condition for one-loop finiteness of the gauge coupling constant
$g$ reads
\be
22\,c_{\rm g}-4\,S_{\rm F}-S_{\rm B}=0\ .
\label{fin8}
\ee
\item Adopting this result, the condition for two-loop finiteness of the
gauge coupling constant~$g$~reads
\be
E(Y)-12\,g^4\,d_{\rm g}\,[Q_{\rm F}+Q_{\rm B}+c_{\rm g}\,(S_{\rm
F}-2\,c_{\rm g})]=0\ ,
\label{fin9}
\ee
with the shorthand notation
\be
E(Y):=6\,g^2\,\tr_{\rm F}\left(C_{\rm F}\sum_{\b=1}^{d_{\rm
B}}Y^{\dg\b}\,Y_\b\right)\ ,
\label{eq:e(y)}
\ee
where by $\tr_{\rm F}$ we indicate the partial trace over the fermionic
indices only, and the abbreviations
\bea
Q_{\rm F}&:=&\sum_If_I\,S_I\,C_I\equiv\frac{1}{d_{\rm g}}\,\tr(C_{\rm F})^2\ ,
\nonumber\\[1ex]
Q_{\rm B}&:=&\sum_Ib_I\,S_I\,C_I\equiv\frac{1}{d_{\rm g}}\,\tr(C_{\rm B})^2\ ,
\label{fin7}
\eea
where the summation index $I$ distinguishes the inequivalent irreducible
representations $R_I$ with multiplicities $f_I$ and $b_I$ in the (completely
reducible) representations $R_{\rm F}$ and $R_{\rm B}$, respectively.
\item The condition for one-loop finiteness of the Yukawa couplings $Y_{\a
ij}$ reads
\bea
&&\sum_{\b=1}^{d_{\rm B}}\left\{4\,Y_\b\,Y^{\dg\a}\,Y_\b+Y_\a\,Y^{\dg\b}\,Y_\b
+Y_\b\,Y^{\dg\b}\,Y_\a+Y_\b\,\tr_{\rm F}\left(Y^{\dg\a}\,Y_\b
+Y^{\dg\b}\,Y_\a\right)\right\}\nonumber\\[1ex]
&&-\;6\,g^2\left[Y_\a\,C_{\rm F}+\tp{\left(C_{\rm F}\right)}\,Y_\a\right]=0\ ;
\label{fin10}
\eea
we call this (cubic and thus troublesome) relation, for brevity, the ``Yukawa
finiteness condition'' (YFC).
\end{itemize}
These three (lowest-order) finiteness conditions for the gauge and Yukawa
couplings, Eqs.~(\ref{fin8}), (\ref{fin9}), and (\ref{fin10}), have been
identified as the central part of the whole set of finiteness conditions: any
investigation of (perturbative) finiteness of quantum field theories should
start from this set of equations \cite{kranner91}. (The quantity $E(Y)$ in
Eq.~(\ref{eq:e(y)}) constitutes the link between the two-loop gauge-coupling
finiteness condition (\ref{fin9}) and the relation one obtains when
multiplying the YFC (\ref{fin10}) by $Y^{\dg\a}$, performing the sum over all
$\a=1,\dots,d_{\rm B}$, and taking the trace of the resulting expression with
respect to the fermionic indices.)

The Yukawa couplings $Y_{\a ij}$ define the Yukawa coupling tensor $Y$, which
has to be invariant under gauge transformations to preserve the gauge
invariance of the theory defined by the Lagrangian (\ref{fin1}):
$$
\left(R_{\rm B}^{\rm c}\otimes R_{\rm F}^{\rm c}\otimes R_{\rm F}^{\rm c}
\right)Y\equiv Y\ ,
$$
where by $R^{\rm c}$ we denote the contragredient of any representation $R$
of $G$. The fermionic representation $R_{\rm F}$ may always be decomposed
into irreducible representations $R_{\rm F}^I$ of $G$, the bosonic
representation $R_{\rm B}$ may always be assumed to be an diagonal alignment
of real orthogonal blocks $R_{\rm B}^\mu$:\footnote{\normalsize\ Any non-real
irreducible representation $R_{\rm B}^A$ which contributes to $R_{\rm B}$ has
to be accompanied by its contragredient counterpart $(R_{\rm B}^A)^{\rm c}$
in order to form the real representation $R_{\rm B}^\mu=(R_{\rm
B}^\mu)^*\simeq R_{\rm B}^A\oplus(R_{\rm B}^A)^{\rm c}$.}
\bea
R_{\rm F}&=&\dirsum{I}\,R_{\rm F}^I\ ,\nonumber\\[1ex]
R_{\rm B}&=&\dirsum{\mu}\,R_{\rm B}^\mu\ .
\label{gau3}
\eea
These direct sums strongly suggest to split every bosonic index $\a$ into a
pair of indices $(A,\a_A)$, where $A$ discriminates the different irreducible
representations $R_{\rm B}^A\subset R_{\rm B}$ and $\a_A\in\{1,\dots,\dim
R_{\rm B}^A\}$ labels the components of $R_{\rm B}^A$, and every fermionic
index $i$ into a pair of indices $(I,i_I)$, where $I$ discriminates the
different irreducible representations $R_{\rm F}^I\subset R_{\rm F}$ and
$i_I\in\{1,\dots,\dim R_{\rm F}^I\}$ labels the components of $R_{\rm F}^I$.
The tensor product $R_{\rm B}^A\otimes R_{\rm F}^I\otimes R_{\rm F}^J$ of
three irreducible representations $R_{\rm B}^A\subset R_{\rm B}$, $R_{\rm
F}^I,R_{\rm F}^J\subset R_{\rm F}$ of $G$ allows for $N(A,I,J)$ invariant
tensors $\Lambda^{(k)}$ if and only if it contains the trivial
representation, $\op{1}$, $N(A,I,J)$ times. In terms of these invariant
tensors, the expansion of $Y$, with coefficients $p^{(k)}_{AIJ}\in\Bbb C$,
reads\footnote{\normalsize\ For a more detailed discussion of the connection
between this expansion of the Yukawa couplings $Y_{\a ij}$ and the
decompositions (\ref{gau3}) of the representations $R_{\rm F}$ and $R_{\rm
B}$, consult Appendix~B of Part~I~\cite{lucha96-1}.}
\be
Y_{\a ij}=Y_{(A,\a_A)(I,i_I)(J,j_J)}=\sum_{k=1}^{N(A,I,J)}p^{(k)}_{AIJ}
\left(\Lambda^{(k)}\right)_{\a_Ai_Ij_J}\ .
\label{gau4}
\ee

For convenience, we define, in terms of the Yukawa couplings $Y_{\a ij}$ and
their Hermitean conjugates, the Hermitean matrix
\be
2\,{x^{i\a}}_{j\b}={\left(Y^{\dg\a}\,Y_\b+Y^{\dg\b}\,Y_\a\right)^i}_j\ ,
\label{gau5}
\ee
which transforms like a gauge-invariant and diagonalizable operator on the
product space $V_{\rm F}\times V_{\rm B}$ \cite{lucha96-1}. Taking partial
traces of this operator $x$ with respect to its bosonic and fermionic
indices, respectively, yields invariant diagonalizable operators $y_{\rm F}$
and $y_{\rm B}$ which commute, of course, with the corresponding Casimir
operators $C_{\rm F}$ and $C_{\rm B}$, respectively. The invariance of the
YFC under arbitrary $U(d_{\rm F})\otimes O(d_{\rm B})$ transformations
\cite{kranner91} enables us to choose $y_{\rm F}$, $C_{\rm F}$, and $y_{\rm
B}$ simultaneously diagonal; in this way, we find, in the fermionic sector,
\bea
\sum_{\b=1}^{d_{\rm B}}{\left(Y^{\dg\b}\,Y_{\b}\right)^i}_j=:
{(y_{\rm F})^i}_j&=&{\d^i}_j\,y_{\rm F}^j\ ,\nonumber\\[1ex]
{(C_{\rm F})^i}_j&=&{\d^i}_j\,C_{\rm F}^j\ ,
\label{gau6f}
\eea
and, in the bosonic sector,
\be
\tr_{\rm F}\left(Y^{\dg\a}\,Y_{\b}+Y^{\dg\b}\,Y_{\a}\right)=:
2\,{(y_{\rm B})^\a}_\b=2\,{\d^\a}_\b\,y_{\rm B}^\b\ .
\label{gau6b}
\ee
We re-order both the fermionic indices $i$ and the bosonic indices $\a$ such
that the first $n\leq d_{\rm F}$ fermionic indices and the first $m\leq
d_{\rm B}$ bosonic indices cover precisely those subsets of $R_{\rm F}$ and
$R_{\rm B}$, respectively, which have non-vanishing Yukawa couplings. This
ordering is then equivalent to the requirement \cite{lucha96-1}
\bea
y_{\rm F}^i\neq 0&\lra&i\in\{1,\dots,n\leq d_{\rm F}\}\ ,\nonumber\\[1ex]
y_{\rm F}^i=0&\lra&i\in\{n+1,\dots,d_{\rm F}\}\ ,\nonumber\\[1ex]
y_{\rm B}^\a\neq 0&\lra&\a\in\{1,\dots,m\leq d_{\rm B}\}\ ,\nonumber\\[1ex]
y_{\rm B}^\a=0&\lra&\a\in\{m+1,\dots,d_{\rm B}\}\ .
\label{gau7}
\eea
By taking into account the diagonalized operators (\ref{gau6f}) and
(\ref{gau6b}), as well as our choice for the ordering of fermionic and
bosonic indices, expressed by Eq.~(\ref{gau7}), the YFC (\ref{fin10}) reduces
to some {\em standard form\/} \cite{lucha96-1,kranner91}:
\be
4\sum_{\b=1}^{m}\left(Y_\b\,Y^{\dg\a}\,Y_\b\right)_{ij}+Y_{\a ij}
\left(2\,y_{\rm B}^\a+y_{\rm F}^i+y_{\rm F}^j
-6\,g^2\,C_{\rm F}^i-6\,g^2\,C_{\rm F}^j\right)=0
\label{gau8}
\ee
for all $i,j\leq n$ and for all $\a\leq m$. This standard form of the YFC is
naturally related to a choice of the bases for the fermionic and bosonic
representation spaces such that both the fermionic representation $R_{\rm F}$
and the bosonic representation $R_{\rm B}$ are blockdiagonal.

As a consequence of the particular structure of the cubic YFC (\ref{fin10}),
the quantity $E(Y)$ introduced by Eq.~(\ref{eq:e(y)}) turns out to be
necessarily bounded:
$$
\sum_{i=1}^{n}\left(y_{\rm F}^i\right)^2\leq
E(Y)=6\,g^2\sum_{i=1}^nC_{\rm F}^i\,y_{\rm F}^i\leq
36\,g^4\sum_{i=1}^n\left(C_{\rm F}^i\right)^2\ .
$$
Imposing the requirement that the actual value of this quantity $E(Y)$
coincides with its upper bound,
$$
E(Y)=36\,g^4\sum_{i=1}^n\left(C_{\rm F}^i\right)^2\ ,
$$
is equivalent to fixing the operator $y_{\rm F}$ to the value $y_{\rm
F}^i=6\,g^2\,C_{\rm F}^i$ for all $i\leq n$. In this case, the~cubic YFC,
Eq.~(\ref{fin10}) or Eq.~(\ref{gau8}), reduces to a (much easier-to-handle)
system of equations merely quadratic in the Yukawa couplings $Y_{\a ij}$:
\bea
\sum_{\b =1}^{m}\left(Y_{\b ij}\,Y_{\b kl}+Y_{\b ik}\,Y_{\b jl}
+Y_{\b il}\,Y_{\b jk}\right)&=&0\quad\forall\ i,j,k,l\in\{1,\dots,n\}\ ,
\nonumber\\[1ex]
\sum_{\a =1}^{m}{\left(Y^{\dg\a}\,Y_\a\right)^i}_j&=&
6\,g^2{\left(C_{\rm F}\right)^i}_j\quad\forall\ i,j\in\{1,\dots,n\}\ ,
\nonumber\\[1ex]
\tr_{\rm F}\left(Y^{\dg\a}\,Y_\b\right)&=&
\tr_{\rm F}\left(Y^{\dg\b}\,Y_\a\right)\quad\forall\ \a,\b\in\{1,\dots,m\}\ .
\label{gau11}
\eea

Following Ref.~\cite{kranner91} (and paralleling the discussion in Part~I
\cite{lucha96-1}), we introduce some quantity called $F^2$ by the definition
\be
F^2:=\frac{E(Y)}{36\,g^4\,d_{\rm g}\,Q_{\rm F}}=
\frac{Q_{\rm F}+Q_{\rm B}+c_{\rm g}\,(S_{\rm F}-2\,c_{\rm g})}{3\,Q_{\rm F}}\ ,
\label{gau9}
\ee
where the second equality holds, of course, only upon application of the
two-loop finiteness condition (\ref{fin9}) for the gauge coupling constant
$g$. This quantity $F^2$ is subject to the requirement $0<F^2\leq 1$
\cite{kranner91}. Numerical investigations \cite{kranner91} revealed that in
all finite quantum field theories the value of $F$ is rather close to
$F^2=1$. This observation led to the conjecture that all finite theories
might be characterized by $F^2=1$. The requirement $F^2=1$ is equivalent to
fixing $n$ in the system (\ref{gau11}) to the value $n=d_{\rm F}$.

We call a quantum field theory ``potentially finite'' if its particle content
fulfills both the finiteness condition (\ref{fin8}) and the inequalities
$0<F^2\leq 1$ for that quantity $F^2$ defined by Eq.~(\ref{gau9}), if the
anomaly index of its fermionic representation, $R_{\rm F}$, vanishes, if its
bosonic representation, $R_{\rm B}$, is real, $R_{\rm B}\simeq R_{\rm B}^*$,
and if, at least, one fundamental invariant tensor, required for the
decomposition (\ref{gau4}) of $Y_{\a ij}$, exists.

\section{($\W$-Fold) Reducibility of the Yukawa Finiteness
Condition}\label{sec:ryfc}

Let us call two arbitrary sets $M_1=\{(R^{\mu_1},R^{I_1},R^{J_1})\}$ and
$M_2=\{(R^{\mu_2},R^{I_2},R^{J_2})\}$ of combinations of real bosonic blocks
$R^{\mu_1},R^{\mu_2}\subset R_{\rm B}$ and irreducible fermionic
representations $R^{I_1},R^{I_2},R^{J_1},R^{J_2}\subset R_{\rm F}$ in
$Y_{(\mu,\a_\mu)(I,i_I)(J,j_J)}$ to be disjoint if and only if
$\{R^{\mu_1}\}\cap\{R^{\mu_2}\}=\{R^{I_1}\}\cap\{R^{I_2}\}=\{R^{J_1}\}\cap
\{R^{J_2}\}=\emptyset$. With this notion of disjointness, we may formulate
\cite{lucha96-1}
\bd
Let $M=\{(R^\mu,R^I,R^J)\mid R^\mu\subset R_{\rm B},\ R^I,R^J\subset R_{\rm
F}\}$ be the set of all combinations of real bosonic blocks and irreducible
fermionic representations in the YFC (\ref{gau4}). If $M$ is the union of
$\W\geq 1$ pairwise disjoint non-empty subsets $M_\w$, $\w=1,2,\dots,\W$,
that is,
$$
M=\bigcup_{\w=1}^\W M_\w\ ,
$$
we call this YFC\/ {\em $\W$-fold reducible}. Rather trivially, an only\/
{\em 1-fold} reducible YFC is called\/ {\em irreducible}.\label{red1}
\ed
As noticed in Part~I~\cite{lucha96-1}, this splitting takes place, for every
index of $Y_{(\mu,\a_\mu)(I,i_I)(J,j_J)}$, with respect to the irreducible
representations in the representation $R_{\rm F}$ and the real blocks in the
representation~$R_{\rm B}$. $\W$-fold reducibility of the YFC entails
corresponding block structures of the operators $x$, $y_{\rm F}$, and $y_{\rm
B}$:
\beax
x&=&\dirsum{\w}\,x_\w\ ,\\[1ex]
y_{\rm F}&=&\dirsum{\w}\,y_{\rm F}^\w\ ,\\[1ex]
y_{\rm B}&=&\dirsum{\w}\,y_{\rm B}^\w\ .
\eeax
Now, every bosonic index $\a=(\mu,\a_\mu)\leq m$ and every fermionic index
$i=(I,i_I)\leq n$ corresponds to a certain irreducible part $M_\w$ of the
YFC. In the following, we shall express this fact by the short-hand notation
$\a\in M_\w$ and $i\in M_\w$, respectively. For example, we write in
all partial traces over fermionic or bosonic indices, respectively,
\beax
y_{\rm F}^i&=&(y_{\rm F}^\w)^i\quad\forall\ i\in M_\w\ ,\\[1ex]
y_{\rm B}^\a&=&(y_{\rm B}^\w)^\a\quad\forall\ \a\in M_\w\ .
\eeax
$\W$-fold reducibility of the YFC splits both the ``fermionic''
representation space $V_{\rm F}$ and the ``bosonic'' representation space
$V_{\rm B}$ into direct sums of subspaces. Each of these subspaces is related
to a certain irreducible component $M_\w$. Clearly, the total fermionic and
bosonic dimensions $n$ and $m$ are obtained by summing over the corresponding
fermionic and bosonic dimensions $n_\w$ and $m_\w$ of these subspaces,
respectively:
\beax
n&=&\sum_{\w=1}^{\W}n_\w\ ,\\[1ex]
m&=&\sum_{\w=1}^{\W}m_\w\ .
\eeax
Accordingly, an $\W$-fold reducible YFC in its standard form (\ref{gau8})
will be decomposed into $\W$ subsystems; each of the $\W$ subsystems will
then possess again the structure of the standard form (\ref{gau8}) of the
YFC:
\be
4\sum_{\b\in M_\w}\left(Y_\b\,Y^{\dg\a}\,Y_\b\right)_{ij}
+Y_{\a ij}\left(2\,y_{\rm B}^\a+y_{\rm F}^i+y_{\rm F}^j-6\,g^2\,C_{\rm F}^i
-6\,g^2\,C_{\rm F}^j\right)=0\quad\forall\ \a, i,j\in M_\w\ .
\label{red5}
\ee
The $\W$ subsystems may be regarded as independent from each other as long as
one does not take into account, for the quantity $E(Y)$, its upper bound
quoted in Eq.~(\ref{gau9}), equivalent to the demand~$F^2\leq 1$. The total
$E(Y)$ is the sum of the individual contributions $E_\w(Y)$ of the
irreducible components $M_\w$:
$$
E(Y)=\sum_{\w=1}^\W E_\w(Y)
=6\,g^2\,\sum_{\w=1}^\W\,\sum_{i\in M_\w}C_{\rm F}^i\,y_{\rm F}^i\ .
$$
Because of the necessity to satisfy that upper bound on $E(Y)$, the
irreducible components $M_\w$ become related to each other and every single
contribution $E_\w(Y)$ is bounded by $E_\w(Y)\leq 36\,g^4\sum_{i\in
M_\w}(C_{\rm F}^i)^2$.

\section{Representations of some Clifford Algebra for a Reducible Yukawa
Finiteness Condition}\label{sec:cred1}

We would like to start our investigations by presenting, irrespective of the
constraints imposed by all the other finiteness conditions, the complete
characterization of a particular subset of solutions~of~an arbitrary
reducible YFC: let's assume that the invariant operator $x$ defined in
Eq.~(\ref{gau5}) is some tensor product of both a bosonic and a fermionic
operator. This particular structure will be shown below to encompass all
Yukawa solutions of the Clifford type. The special case of an irreducible YFC
has been covered in Part~I~\cite{lucha96-1}. We may transfer the results
obtained there to every single irreducible component of the general YFC under
consideration here---which simplifies the present analysis considerably. The
precise formulation of our assumption concerning $x$ is given in
\bd
Let $M=\{(R^\mu,R^I,R^J)\mid R^\mu\subset R_{\rm B},\ R^I,R^J\subset R_{\rm
F}\}$ be the set of all combinations of real orthogonal bosonic blocks and
irreducible fermionic representations in an $\W$-fold reducible YFC:
$$
M=\bigcup_{\w=1}^\W M_\w\ .
$$
For each irreducible component $M_\w$, we assume the operator $x$, defined by
Eq.~(\ref{gau5}), to be of the~form\footnote{\normalsize\ This ansatz is
motivated by the observation that the irreducible components $M_\w$ of the
YFC~(\ref{gau8}) define some decomposition of the fermionic and bosonic
representation spaces $V_{\rm F}$ and $V_{\rm B}$, respectively, into
subspaces corresponding to the particular irreducible components $M_\w$,
which allows to introduce operators $u_\w$ and $v_\w$ acting on these
subspaces.}
$$
x_\w=u_\w\otimes v_\w\ ,
$$
with the factor $u_\w$ carrying only
bosonic indices and the factor $v_\w$ carrying only fermionic
indices.\label{cred2}
\ed
Under this assumption, every single irreducible component $x_\w$ of $x$ may
be shown to be of the form\footnote{\normalsize\ NB: the sum in the
denominator of $x^{i\a}$ involves only $k\in M_\w$; in the case of an
irreducible YFC, this sum extends over all $i\leq n$.}
\be
{x^{i\a}}_{j\b}={(x_\w)^{i\a}}_{j\b}={\d^\a}_\b\,{\d^i}_j\,x_\w^{j\b}=
{\d^\a}_\b\,{\d^i}_j\,\frac{y_{\rm B}^\b\,y_{\rm F}^j}
{\ds\sum_{k\in M_\w}y_{\rm F}^k}\quad\forall\ i,j,\a,\b\in M_\w\ .
\label{cred3}
\ee
As a consequence of our choice (\ref{gau7}) for the ordering of fermionic and
bosonic indices, every irreducible component $x_\w$ is invertible. The tensor
structure ansatz for $x_\w$ in Def.~\ref{cred2} enforces some block structure
of $Y_\a$, controlled by $y_{\rm F}^i$ \cite{lucha96-1}:
\be
Y_{\a ij}\left(y_{\rm F}^i-y_{\rm F}^j\right)=0\quad\forall\ \a,i,j\in M_\w\ .
\label{cred4}
\ee
Now, with the help of the definition~(\ref{cred3}) of $x_\w^{i\a}$ and the
commutator (\ref{cred4}), every irreducible subsystem of the YFC may be cast
from the standard form (\ref{red5}) to a form quasi-linear in the Yukawa
couplings:
\be
Y_{\a ij}\left(8\,x_\w^{i\a}-2\,y_{\rm F}^i+2\,y_{\rm B}^\a-
6\,g^2\,C_{\rm F}^i-6\,g^2\,C_{\rm F}^j\right)=0\quad\forall\ \a,i,j\in M_\w\ .
\label{cred5}
\ee
\br
{\em The restriction of Eq.~(\ref{cred5}) to the value $F^2=1$ (by demanding
$y_{\rm F}^i=6\,g^2\,C_{\rm F}^i$ and $n=d_{\rm F}$) implies, for every
irreducible component $M_\w$ of the YFC, the constraints $4+n_\w=2\,m_\w$ and
$y_{\rm F}^i=y_\w$ for all $i\in\{1,\dots,n_\w\}$.}\label{cred6}
\er

In order to solve the YFC in the quasi-linear form (\ref{cred5}), we note
that the only quantity there which does not depend on the Yukawa couplings
$Y$ is the expression $6\,g^2\,C_{\rm F}$, which is also independent~of~$\a$,
and that this relation is linear in $x^{i\a}$. This implies that, for this
set of equations to be solvable at~all, the quantities $x_\w^{i\a}$ must be
of the order $\mbox{O}(g^2)$. Therefore, a reasonable {\em ansatz\/} for
solving the YFC~(\ref{cred5}) in each of its irreducible components $M_\w$
reads
\be
x_\w^{i\a}=6\,g^2\,a_\w\,C_{\rm F}^i+b_\w\quad\forall\ i,\a\in M_\w\ ,
\label{cred7}
\ee
involving the two arbitrary constants $a_\w,b_\w\in{\Bbb C}$. The procedure
for solving the quasi-linear YFC~(\ref{cred5}) by this ansatz has already
been sketched in Part~I~\cite{lucha96-1}. In the case $a_\w\neq 0$, the
commutator (\ref{cred4}) entails
\be
Y_{\a ij}\left(C_{\rm F}^i-C_{\rm F}^j\right)=0\ .
\label{cred8}
\ee
This commutator of Yukawa couplings and fermionic Casimir operator may then
be used to formulate
\bp
Let the YFC be $\W$-fold reducible; let for all irreducible components
$M_\w$, $\w=1,\dots,\W$, the Yukawa couplings be compatible with both the
tensor structure assumption $x_\w=u_\w\otimes v_\w$ of Def.~\ref{cred2} and
the ansatz $x_\w^{i\a}=6\,g^2\,a_\w\,C_{\rm F}^i+b_\w$ of Eq.~(\ref{cred7}).
Then, for every irreducible system (\ref{cred5}), the solutions for $y_{\rm
F}^i$ necessarily assume one of the following values:
\begin{enumerate}
\item[A:] For $a_\w=0$, only one common value for all $y_{\rm F}^i$,
proportional to some average $C_{\rm m}^\w:=(C_{\rm F}^i+C_{\rm F}^j)/2$ of
Casimir eigenvalues, is conceivable:
$$
y_{\rm F}^i\equiv y_\w=6\,g^2\,\frac{m_\w}{4-m_\w+n_\w}\,C_{\rm m}^\w\quad
\forall\ i\in M_\w\ .
$$
\item[B:] For $a_\w\neq(4-m_\w)^{-1}$, only one fermionic Casimir eigenvalue
$C_\w$ is admissible, ${(C_{\rm F})^i}_j={\d^i}_j\,C_\w$, and only one common
value $y_\w$ for all $y_{\rm F}^i$ is allowed:
$$
y_{\rm F}^i\equiv y_\w=6\,g^2\,\frac{m_\w}{4-m_\w+n_\w}\,C_\w\quad\forall\
i\in M_\w\ .
$$
\item[C:] For $a_\w=(4-m_\w)^{-1}$, different values for $y_{\rm F}^i$ are
possible:
$$
(4-m_\w)\,y_{\rm F}^i=6\,g^2\,m_\w\left(C_{\rm F}^i-\frac{1}{4-m_\w+n_\w}
\sum_{k\in M_\w}C_{\rm F}^k\right)\quad\forall i\in M_\w\ .
$$
\end{enumerate}\label{cred9}
\ep
\br
{\em Proposition~\ref{cred9} reveals that it is not obligatory to demand
$F^2=1$ in order to find solutions of the YFC and that, according to Case C,
even solutions with different values of $y_{\rm F}^i$ can be
obtained.}\label{cred10}
\er

By application of a transformation of the kind employed in Appendix~D of
Part~I~\cite{lucha96-1}, the Yukawa couplings satisfying the particular
ansatz in Def.~\ref{cred2} can be shown, for each irreducible component
$M_\w$, to be equivalent to generators of the representation of some Clifford
algebras with identity element~$\op{1}$; $m_\w-1$ Yukawa couplings become
equivalent to real and symmetric generators $N_\a$, $\a=1,\dots,m_\w-1$, of
these Clifford algebra representations, forming the subset of generators
$$
\fb_\w=\{N_\a\mid\akomm{N_\a}{N_\b}=2\,\d_{\a\b}\,\op{1}_{n_\w},\
N_{\a ij}=N_{\a ji}\in{\Bbb R},\ \a=1,\dots,m_\w-1\}\ .
$$
Re-phrasing part of Remark \ref{cred10}, we note that the requirement $F^2=1$
is not necessary in order to find Clifford-type solutions of the YFC, as has
already been shown for the case of an irreducible YFC \cite{lucha96-1}.
\br
{\em For a fixed irreducible component $M_\w$ of the YFC, the involved
Clifford algebras and the subset ${\fb}_\w$ of their representations can be
chosen independently from all other irreducible components of the YFC.}
\er
The above Clifford algebra structure restricts, for each irreducible
component $M_\w$, the possible ranges of the effective bosonic and fermionic
dimensions $m_\w$ and $n_\w$. The rank $p_i$ of some Clifford
algebra~${\fc}_i$ is either even, $p_i=2\,\nu_i$, or odd, $p_i=2\,\nu_i+1$,
with $\nu_i\in\Bbb N$. In both cases, every matrix representation of
${\fc}_i$ is built from $2^{\nu_i}$-dimensional blocks
\cite{boerner55,riesz93}. This implies an inequality for the dimension of the
particular representation and the number of symmetric generators of ${\fc}_i$
realized as Yukawa couplings:
\be
n_\w\geq 2^{m_\w-2}\ .
\label{cred12}
\ee

We are thus in the position to search, with the help of numerical computer
methods, for potentially finite quantum field theories with Yukawa couplings
of the Clifford type. All potentially finite theories for a given simple Lie
algebra $\fa$ are provided by a (recently developed) C package
\cite{package}, which contains additionally an optional user-defined function
{\tt constraint}, to be employed to perform further checks. Any theory
passing this {\tt constraint} is characterized by a fermionic multiplicity
vector $\mb{f}\equiv(f_I)$ and a bosonic multiplicity vector
$\mb{b}\equiv(b_A)$, whose components are the multipicities of mutually
inequivalent irreducible representations $R^I\in R_{\rm F}$ and $R^A\in
R_{\rm B}$, respectively. In all analyses to be presented below, we denote
a specific $d$-dimensional irreducible representation of some Lie algebra
$\fa$ by the symbol~$[d]$.

\section{Representations of Clifford Algebras for $F^2=1$
Theories?}\label{sec:rca}

First, we employ the subroutine {\tt constraint} of our C package in order
to implement the requirements $F^2=1$ and, in agreement with
Remark~\ref{cred6},
$$
n=\sum_{\w=1}^{\W}n_\w=d_{\rm F}\ .
$$
The bosonic and fermionic dimensions of every irreducible component $M_\w$ of
the YFC are related by
$$
4+n_\w=2\,m_\w\ ,
$$
indicating that any fermionic dimension $n_\w$ must be even. This relation
may then be used to eliminate the bosonic dimension $m_\w$ from the
inequality (\ref{cred12}), with the---for the following rather
crucial---result
$$
n_\w\geq 2^{m_\w-2}=2^{n_\w/2}\ ,
$$
from which we deduce that the fermionic dimension $n_\w$ is necessarily
restricted to one of three values: $n_\w=2,3,4$. Consequently, for
potentially finite $F^2=1$ theories with Clifford-type Yukawa couplings,
there are no more than two options for the dimensions of any irreducible
component $M_\w$ of the YFC: $(n_\w=2,m_\w=3)$ or $(n_\w=4,m_\w=4)$. Needless
to say, every irreducible component $M_\w$ of the YFC has to embrace both
complete irreducible fermionic representations $R_{\rm F}^I$ and complete
real orthogonal bosonic blocks $R_{\rm B}^\mu$ of representations of the Lie
algebra $\fa$, coupling invariantly within the component $M_\w$. Direct
inspection of all simple Lie algebras $\fa$ shows that only the four Lie
algebras~$A_1$,~$A_2$,~$A_3$, and $B_2$ possess irreducible representations
of sufficiently low dimension for use in the fermionic sector. Case-by-case
examination of all potentially finite theories extracted in this manner
allows us to state
\bp
There are no potentially finite $F^2=1$ solutions of the quasi-linear
irreducible system (\ref{cred5}) obeying simultaneously the relation
(\ref{cred12}) for the corresponding bosonic and fermionic dimensions.
\ep
The detailed (and straightforward) line of reasoning leading to this result
can be found in Appendix~\ref{app:pff=1tcyc}.

\section{Representations of Clifford Algebras for General
Theories}\label{sec:rcagt}

With enhanced confidence in our formalism, by feeding the algebraic
expressions given in Prop.~\ref{cred9}~into the user-defined function {\tt
constraint} of our C-package \cite{package}, we proceed to investigate the
general~case of theories {\em not\/} constrained by $F^2=1$. For every
potentially finite theory provided by this subroutine, both Inequality
(\ref{cred12}) as well as all the hypothetical values of $y_{\rm F}$ allowed
by Prop.~\ref{cred9} have to be checked. For simplicity, we confine our
investigations to those theories where all irreducible representations~in
$R_{\rm F}$ and $R_{\rm B}$ which may form gauge-invariant Yukawa couplings
in the sense of Eq.~(\ref{gau4}) indeed couple. Moreover, computer resources
force us to focus to the case $a\neq 0$. First, any potentially finite theory
is characterized by a fermionic multiplicity vector $\mb{f}^0=(f_I^0)$ and a
bosonic multiplicity~vector~$\mb{b}^0=(b_\mu^0)$:
\beax
R_{\rm F}&=&\dirsum{I}\,f_I^0\,R^I\ ,\\[1ex]
R_{\rm B}&=&\dirsum{\mu}\,b_\mu^0\,R^\mu\ .
\eeax
In principle, these representations $R_{\rm F}$ and $R_{\rm B}$ may contain
some irreducible representations of the Lie algebra $\fa$ which have no
appropriate partners to form an invariant coupling obeying the commutator
(\ref{cred8}). Precisely for this reason, we introduce new ``effective''
multiplicity vectors $\mb{f}=(f_I)$ and $\mb{b}=(b_\mu)$ which contain only
those irreducible representations in $R_{\rm F}$ and $R_{\rm B}$,
respectively, which contribute~to Yukawa couplings compatible with the block
structure (\ref{cred4}). These effective multiplicity vectors define the
``YFC-relevant'' fermionic and bosonic representations $R_{\rm F}^{\rm YFC}$
and $R_{\rm B}^{\rm YFC}$:
\bea
R_{\rm F}^{\rm YFC}&=&\dirsum{I}\,f_I\,R^I\subset R_{\rm F}\ ,\nonumber\\[1ex]
R_{\rm B}^{\rm YFC}&=&\dirsum{\mu}\,b_\mu\,R^\mu\subset R_{\rm B}\ .
\label{numerics2}
\eea
All irreducible representations $R^I\subset R_{\rm F}^{\rm YFC}$ and
$R^\mu\subset R_{\rm B}^{\rm YFC}$ find suitable partners to build invariants:
$$
R^I\otimes R^J\otimes R^\mu\supset\op{1}\quad\mbox{for}\
R^I\subset R_{\rm F}^{\rm YFC},\
R^J\subset R_{\rm F}^{\rm YFC},\
R^\mu\subset R_{\rm B}^{\rm YFC}\ .
$$
\newpage\no
Expressed in terms of the dimensions $d_I:=\dim R^I$ of the irreducible
fermion representations $R^I$ and the dimensions $d_\mu:=\dim R^\mu$ of the
real orthogonal boson representations $R^\mu$, respectively, the~total
fermionic and bosonic dimensions $n$ and $m$ then read
\beax
n=\dim R_{\rm F}^{\rm YFC}&=&\sum_If_I\,d_I\leq d_{\rm F}\ ,\\[1ex]
m=\dim R_{\rm B}^{\rm YFC}&=&\sum_\mu b_\mu\,d_\mu\leq d_{\rm B}\ .
\eeax
Introducing the fermionic multiplicity vector $\mb{f}_\w=(f_I^\w)$ and the
bosonic multiplicity vector $\mb{b}_\w=(b_\mu^\w)$ corresponding to a
particular irreducible component $M_\w$ of an $\W$-fold reducible YFC, any
distribution of {\em all\/} irreducible representations in
Eq.~(\ref{numerics2}) on the different subsystems $M_\w$ yields the
decompositions
\bea
R_{\rm F}^{\rm YFC}&=&\sum_{\w=1}^\W R_{\rm F}^\w=
\sum_{\w=1}^\W\,\dirsum{I}\,f_I^\w\,R^I\ ,\nonumber\\[1ex]
R_{\rm B}^{\rm YFC}&=&\sum_{\w=1}^\W R_{\rm B}^\w=
\sum_{\w=1}^\W\,\dirsum{\mu}\,b_\mu^\w\,R^\mu\ ,
\label{numerics5}
\eea
The bosonic dimension $m_\w$ and the fermionic dimension $n_\w$ of an
irreducible component $M_\w$~thus~read
\beax
n_w&=&\sum_If_I^\w\,d_I\ ,\\[1ex]
m_w&=&\sum_\mu b_\mu^\w\,d_\mu\ .
\eeax
According to our previous assumptions, we must guarantee that all irreducible
representations which are able to build invariant tensors actually
contribute:\footnote{\normalsize\ In general, the artificial splitting of the
YFC into irreducible subsystems could destroy the ability of an irreducible
representation to find partners for invariant Yukawa couplings. This is due
to Def.~\ref{red1} which demands that any irreducible representation
belonging to an irreducible component $M_\w$ of the YFC must find its
partners to form invariant couplings within the same irreducible component
$M_\w$.}
\bea
f_I=\sum_{\w=1}^\W f_I^\w\ ,\nonumber\\[1ex]
b_\mu=\sum_{\w=1}^\W b_\mu^\w\ .
\label{numerics7}
\eea
Any such distribution of irreducible representations consistent with
Eqs.~(\ref{numerics5}) to (\ref{numerics7}) has to
be~found.\footnote{\normalsize\ Algorithms like this must be recursive: every
$f_I^\w$ runs from~0~to~$f_I$ and every $b_\mu^\w$ runs from~0~to~$b_\mu$.}

For every combination of multiplicity vectors $\mb{f}_\w$ and $\mb{b}_\w$
allowing to construct invariant tensors~in the respective irreducible
component $M_\w$, the value of $F^2$ resulting from these YFC-relevant
fermionic and bosonic multiplicities has to be compared with the value
(\ref{gau9}) of $F^2$ determined by the full particle content of the
potentially finite theory considered. Furthermore, for every irreducible
component $M_\w$, Inequality (\ref{cred12}), necessary for every construction
of Clifford-type Yukawa couplings, has to be checked.

The calculation of $F^2_{\rm YFC}$ is straightforward \cite{lucha96-1}: The
number of different Casimir eigenvalues $C_I$ in $M_\w$ tells us whether
Case~B or Case~C of Prop.~\ref{cred9} is realized in this $M_\w$. Introducing
the abbreviations
\beax
Q_{\rm F}&=&\sum_{R^I\subset R_{\rm F}}f_I^0\,S_I\,C_I\ ,\\[1ex]
Q^\w&=&\sum_{R^I\subset R_{\rm F}^\w}f_I^\w\,S_I\,C_I\ ,\\[1ex]
S^\w&=&\sum_{R^I\subset R_{\rm F}^\w}f_I^\w\,S_I\ ,\\[1ex]
C_0^\w&=&\frac{1}{4-m_\w+n_\w}\sum_{R^I\subset R_{\rm F}^\w}f_I^\w\,d_I\,C_I=
\frac{S^\w\,d_{\rm g}}{4-m_\w+n_\w}\ ,
\eeax
we immediately find for the individual contribution $F^2_\w$ of this $M_\w$
to $F^2$, if $C_{\rm F}^I=C_\w$ for all $f_I^\w\neq 0$,
$$
F^2_\w=\frac{m_\w}{4-m_\w+n_\w}\,\frac{Q^\w}{Q_{\rm F}}\ ,
$$
else
$$
F^2_\w=\frac{m_\w}{4-m_\w}\,\frac{Q^\w-C_0^\w\,S^\w}{Q_{\rm F}}\ .
$$
The total value of $F^2$ is then obtained as the sum of the individual
contributions $F^2_\w$ of the irreducible components $M_\w$:
$$
F^2_{\rm YFC}=\sum_{\w=1}^\W F^2_\w\ .
$$
The question we must answer for every candidate theory is: Are any of the
Yukawa solutions obtained in our Prop.~\ref{cred9} compatible with the
two-loop gauge-coupling finiteness condition (\ref{fin9})? In other words, is
$F^2_{\rm YFC}$---corresponding to the defining expression in
Eq.~(\ref{gau9})---identical to the value of $F^2$ arising~from the particle
content of the theory---the right-hand side of Eq.~(\ref{gau9}):
$$
F^2_{{\rm YFC}}=\frac{Q_{\rm F}+Q_{\rm B}+c_{\rm g}\,(S_{\rm F}-2\,c_{\rm g})}
{3\,Q_{\rm F}}\ ?
$$
Every theory passing this criterion and satisfying Inequality (\ref{cred12})
may be regarded as a good candidate for a finite quantum field theory with
Clifford-like Yukawa couplings.

Remarkably, the numerical scan with the help of our C package \cite{package}
through all simple Lie algebras ${\fa}=(A_r,B_r,C_r,D_r,E_6,E_7,E_8,F_4,
G_2)$ up to $r=\mbox{rank}\,{\fa}\leq 8$ does not produce any viable
candidate:
\bp
For $x_\w=u_\w\otimes v_\w$ as in Def.~\ref{cred2} and the ansatz
(\ref{cred7}) for $x^{i\a}$, there does not exist~any potentially finite
theory with Yukawa couplings satisfying an $\W$-fold reducible YFC, at least
for simple gauge groups with rank less than or equal to 8, if all irreducible
representations allowing for invariant tensors for the Yukawa couplings
really contribute.
\ep

\section{Summary, Conclusions, and Outlook}\label{sec:sco}

Combining our findings presented in Part~I \cite{lucha96-1} with the results
obtained here, our investigation of~the relevance of Clifford algebras for
the construction of Yukawa couplings in finite quantum field~theories may be
summarized in form of two theorems. For $F^2=1$, our analysis is totally
general and complete:
\bt
Let the YFC be $\W$-fold reducible, and assume for the Yukawa couplings the
Clifford~form $x_\w=u_\w\otimes v_\w$, $1\leq\w\leq\W$; then there do not
exist $F^2=1$ solutions of the YFC satisfying the following criteria:
\begin{enumerate}
\item The fermionic representation $R_{\rm F}$ has vanishing anomaly index.
\item The bosonic representation $R_{\rm B}$ is real (orthogonal).
\item The beta function for the gauge coupling $g$ vanishes in one-loop
approximation.
\end{enumerate}
\et
Dropping the constraint $F^2=1$, we have to introduce additional (but rather
reasonable) assumptions:
\bt
Let us consider an arbitrary simple Lie algebra\/ $\fa$ with rank less than or
equal to 8,~i.e.,
$$
{\fa}\in
\{\mbox{A}_r,\mbox{B}_r,\mbox{C}_r,\mbox{D}_r,\mbox{E}_6,\mbox{E}_7,
\mbox{E}_8,\mbox{F}_4,\mbox{G}_2\mid r=\mbox{rank}\,{\fa}\leq 8\}\ ,
$$
let the YFC be $\W$-fold reducible, and assume for the Yukawa couplings the
Clifford form $x_\w=u_\w\otimes v_\w$, $1\leq\w\leq\W$; then there does not
exist any solution of the YFC with a genuine linear dependence of~$x^{i\a}$
on the fermionic Casimir eigenvalues $C_{\rm F}^i$, i.e.,
$$
x^{i\a}=6\,g^2\,a_\w\,C_{\rm F}^i+b_\w\quad ,\ a_\w,b_\w\in{\Bbb C}\quad ,\
a_\w\neq 0\ ,
$$
satisfying the following criteria:
\begin{enumerate}
\item The fermionic representation $R_{\rm F}$ has vanishing anomaly index.
\item The bosonic representation $R_{\rm B}$ is real (orthogonal).
\item The beta function for the gauge coupling $g$ vanishes in one- and
two-loop approximation.
\newpage
\item Irreducible blocks $R_{\rm B}^\mu\subset R_{\rm B}$ and $R_{\rm
F}^I,R_{\rm F}^J\subset R_{\rm F}$, with multiplicities $b_\mu$, $f_I$, and
$f_J$, respectively, which allow for invariant couplings, i.e., $R_{\rm
B}^\mu\otimes R_{\rm F}^I\otimes R_{\rm F}^J\supset\op{1}$, contribute to the
YFC such that
$$
\left.y_{\rm F}\right|_{\op{1}_{f_I}\times R^I}\neq 0
$$
and
$$
\left.y_{\rm B}\right|_{\op{1}_{b_\mu}\times R^\mu}\neq 0\ .
$$
\end{enumerate}
\et
Accordingly, the chances for Clifford algebras in finite theories are not too
good. Nevertheless, in~order to promote the understanding of the algebraic
structure(s) of the conditions for finiteness of arbitrary quantum field
theories, we shall discuss in Ref.~\cite{lucha96-2} the true origin of this
Clifford-algebra conjecture.

\section*{Acknowledgements}

M.~M.~was supported by the ``Fonds zur F\"orderung der wissenschaftlichen
Forschung in \"Osterreich,'' project 09872-PHY, by the Institute for High
Energy Physics of the Austrian Academy of Sciences, and by a grant of the
University of Vienna.

\newpage

\appendix

\section{Potentially Finite $F^2=1$ Theories Involving Clifford-Like Yukawa
Couplings}\label{app:pff=1tcyc}

\subsection{The Lie algebra A$_1$}

For the Lie algebra A$_1$, the only irreducible representations of dimensions
less than or equal to 4 are the two-, three-, and four-dimensional
representations [2], [3], [4]. All potentially finite $F^2=1$ theories based
on A$_1$ with fermionic representations $R_{\rm F}$ containing only these
three irreducible representations are listed (consecutively numbered) in
Table~\ref{tab:a1}. The appearance of (any number of) three-dimensional
irreducible representations in the fermionic representation $R_{\rm F}$ of a
potentially finite theory is certainly incompatible with either of the two
conceivable values, $n_\w=2$ or $n_\w=4$, of the fermionic dimension $n_\w$
of any irreducible component $M_\w$ of our YFC. Inspection of
Table~\ref{tab:a1} leaves us with two candidates:
\begin{itemize}
\item Theory no.~1 has already been extensively discussed, and subsequently
ruled out, in Part~I~\cite{lucha96-1}.
\item Theory no.~4 involves the four-dimensional irreducible fermion
representation [4], to be covered by an irreducible component $M_\w$ of
fermionic dimension $n_\w=4$, which, in turn, implies $m_\w=4$ for its
bosonic dimension. However, since $[4]\otimes[4]\not\supset[2]$, there is no
suitable invariant tensor~$\Lambda^{(k)}$.
\end{itemize}
Thus there remains no candidate for a Clifford-type finite $F^2=1$ theory
based on the Lie algebra~A$_1$.

\begin{table}[htb]
\caption{Potentially finite $F^2=1$ theories for the Lie algebra A$_1$ with
fermionic representations $R_{\rm F}$ involving only irreducible
representations of dimension less than or equal to 4. The multiplicities of a
$d$-dimensional irreducible representation of A$_1$ in $R_{\rm F}$ and
$R_{\rm B}$ are denoted by $f_{[d]}$ and $b_{[d]}$,
respectively.}\label{tab:a1}
\begin{center}
\begin{tabular}{rrrrrrr}
\hline\hline\\[-1.5ex]
Theory no.&$\qquad f_{[2]}$&$f_{[3]}$&$f_{[4]}$&$\qquad
b_{[2]}$&$b_{[3]}$&$b_{[4]}$\\[1ex]
\hline\\[-1.5ex]
1&0&0&1&20&7&0\\
2&0&3&0&32&2&0\\
3&0&4&0&0&6&0\\
4&2&0&1&8&8&0\\
5&2&1&1&4&0&2\\
6&2&3&0&20&3&0\\
7&4&2&0&40&0&0\\
8&4&3&0&8&4&0\\
9&6&2&0&28&1&0\\
10&8&2&0&16&2&0\\
11&10&2&0&4&3&0\\
12&12&1&0&24&0&0\\
13&14&1&0&12&1&0\\
14&16&1&0&0&2&0\\[1ex]
\hline\hline
\end{tabular}
\end{center}
\end{table}

\subsection{The Lie algebra A$_2$}

For the Lie algebra A$_2$, the only irreducible representation of dimension
less than or equal to 4 is the three-dimensional fundamental representation
[3]. Obviously, it is not possible to construct invariant $n_\w=2$ or
$n_\w=4$ blocks from three-dimensional representations only. This fact rules
out any theory based on A$_2$.

\subsection{The Lie algebra A$_3$}

For the Lie algebra A$_3$, the only irreducible representation of dimension
less than or equal to 4 is the (four-dimensional) fundamental representation
[4]. However, there exists no potentially finite $F^2=1$ theory with a
fermionic representation $R_{\rm F}$ which involves only this representation
[4]; in other words, every fermionic representation $R_{\rm F}$ in
potentially finite $F^2=1$ theories based on A$_3$ contains at least one
irreducible representation of dimension greater than 4. This circumstance
rules out every theory based on A$_3$.

\subsection{The Lie algebra B$_2$}

For the Lie algebra B$_2$, the only irreducible representation of dimension
less than or equal to 4 is the four-dimensional fundamental representation,
[4]. All potentially finite $F^2=1$ theories based on B$_2$ with fermionic
representations $R_{\rm F}$ involving only this irreducible representation
are listed in Table~\ref{tab:b2}; all corresponding bosonic representations
$R_{\rm B}$ involve the four-, five-, and ten-dimensional irreducible
representations [4], [5], [10] of B$_2$. Every four-dimensional irreducible
representation in the fermionic representation of any of these candidate
theories must be covered by an irreducible component $M_\w$ of fermionic
dimension $n_\w=4$. This, in turn, fixes the bosonic dimension of this
particular irreducible component $M_\w$ to the value $m_\w=4$. However, the
tensor product of two four-dimensional irreducible representations [4] does
not contain the four-dimensional irreducible representation [4]:
$[4]\otimes[4]\not\supset[4]$. Consequently, no appropriate gauge-invariant
tensors may be constructed. We conclude that the Lie algebra B$_2$ provides
no viable candidate for a finite $F^2=1$ theory with Clifford-like Yukawa
couplings.

\begin{table}[htb]
\caption{Potentially finite $F^2=1$ theories for the Lie algebra B$_2$ with
fermionic representations $R_{\rm F}$ involving only irreducible
representations of dimension less than or equal to 4. The multiplicities of a
$d$-dimensional irreducible representation of B$_2$ in $R_{\rm F}$ and
$R_{\rm B}$ are denoted by $f_{[d]}$ and $b_{[d]}$,
respectively.}\label{tab:b2}
\begin{center}
\begin{tabular}{rrrrr}
\hline\hline\\[-1.5ex]
Theory no.&$\qquad f_{[4]}$&$\qquad b_{[4]}$&$b_{[5]}$&$b_{[10]}$\\[1ex]
\hline\\[-1.5ex]
1&29&14&1&0\\
2&30&4&4&0\\
3&31&2&0&1\\[1ex]
\hline\hline
\end{tabular}
\end{center}
\end{table}


\begin{thebibliography}{30}
\bibitem{lucha96-1}
W.~Lucha and M.~Moser, {\em Clifford algebras in finite quantum field
theories, I: Irreducible Yukawa finiteness condition}, HEPHY-PUB 653/96,
UWThPh-1996-49, {\bf hep-th/9702137}; Int.~J.~Mod.\ Phys.\ (in press).
\bibitem{lucha86-1}
W.~Lucha and H.~Neufeld, Phys.~Rev.~D {\bf 34} (1986) 1089.
\bibitem{lucha86-2}
W.~Lucha and H.~Neufeld, Phys.~Lett.~B {\bf 174} (1986) 186.
\bibitem{boehm87}
M.~B\"ohm and A.~Denner, Nucl.~Phys.~B {\bf 282} (1987) 206.
\bibitem{lucha87-1}
W.~Lucha and H.~Neufeld, Helvetica Physica Acta {\bf 60} (1987) 699.
\bibitem{qdiv1}
W.~Lucha, Phys.~Lett.~B {\bf 191} (1987) 404.
\bibitem{qdiv2}
W.~Lucha and M.~Moser, Int.~J.~Mod.~Phys.\ {\bf A9} (1994) 2773.
\bibitem{kranner91}
G.~Kranner and W.~Kummer, Phys.~Lett.~B {\bf 259} (1991) 84.
\bibitem{llewellyn73}
C.~H.~Llewellyn Smith, Phys.~Lett.~B {\bf 46} (1973) 233;\\
J.~M.~Cornwall, D.~N.~Levin, and G.~Tiktopoulos, Phys.~Rev.~D {\bf 10} (1974)
1145.
\bibitem{cheng74}
T.~P.~Cheng, E.~Eichten, and L.-F.~Li, Phys.~Rev.~D {\bf 9} (1974) 2259;\\
M.~E.~Machacek and M.~T.~Vaughn, Nucl.~Phys.~B {\bf 222} (1983) 83.
\bibitem{boerner55}
H.~Boerner, {\em Darstellungen von Gruppen\/} (Springer, Berlin --
Heidelberg, 1967).
\bibitem{riesz93}
M.~Riesz, {\em Clifford Numbers and Spinors}, in {\em Fundamental Theories of
Physics}, E.~F.~Bolinder and P.~Lounesto, eds., (Kluwer Academic Pubs.,
Dordrecht, 1993).
\bibitem{package}
W.~Lucha and M.~Moser, {\em FINBASE, a C package for potentially finite
quantum field theories}, HEPHY-PUB 656/96, UWThPh-1996-52 (in preparation).
This C package will be available at {\tt
http://doppler.thp.univie.ac.at/}\~{}{\tt mmoser}.
\bibitem{lucha96-2}
W.~Lucha and M.~Moser, {\em Clifford algebras in finite quantum field
theories, III: Origin of Clifford solutions}, HEPHY-PUB 655/96,
UWThPh-1996-51 (in preparation).
\end{thebibliography}
\end{document}